\documentclass[11pt,reqno]{amsart}
\usepackage{t1enc}
\usepackage[latin1]{inputenc}
\usepackage{amsmath,latexsym,amssymb,graphicx,dsfont,amsthm,amsfonts}
\usepackage{color}
\usepackage{umoline}
\usepackage{float}
\usepackage{apalike}

\renewcommand{\O}{\mathbf{O}}

\newcommand{\elo}[1]{\mathrm{Elo}_{#1}}

\begin{document}

\numberwithin{equation}{section}

\title[Prediction Model for the Africa Cup of Nations 2019 via Nested Poisson Regression]{Prediction Model for the Africa Cup of Nations 2019 via Nested Poisson Regression}
\author{Lorenz A. Gilch}


\address{Lorenz A. Gilch: Universit\"at Passau, Innstrasse 33, 94032 Passau, Germany}

\email{Lorenz.Gilch@uni-passau.de}
\urladdr{http://www.math.tugraz.at/$\sim$gilch/}
\date{\today}
\keywords{Africa Cup of Nations 2019; football; Poisson regression; Elo}

\maketitle

\begin{abstract}
This article is devoted to the forecast of the Africa Cup of Nations 2019 football tournament. It is based on a Poisson regression model that includes the Elo points of the participating teams as covariates and incorporates differences of team-specific skills. The proposed model allows predictions in terms of probabilities in order to quantify the chances for each team to reach a certain stage of the tournament. Monte Carlo simulations are used to estimate the outcome of each single match of the tournament and hence to simulate the whole tournament itself. The model is fitted on all football games on neutral ground of the participating teams since 2010. 
\end{abstract}

\section{Introduction}

\subsection{Problem formulation}

Football is a typical low-scoring game and games are frequently decided through single events in the game. These events may be extraordinary individual performances,  individual errors, injuries,  refereeing errors or just lucky coincidences.  Moreover, during a tournament there are most of the time teams and players that are in exceptional shape and have a strong influence on the outcome of the tournament. One consequence is that every now and then alleged underdogs win tournaments and reputed favorites drop out already in the group phase.

The above  effects are notoriously difficult to forecast.
Despite this fact, every team has its strengths and weaknesses (e.g., defense and attack) and most of the results reflect the qualities of the teams.  In order to model the random effects and the ``deterministic'' drift  forecasts should be given in terms of  probabilities.

Among football experts and fans alike there is mostly a consensus on the top favorites, e.g. Senegal, Cameroon or Egypt, and more debate on possible underdogs. However, most of these predictions rely on subjective opinions and are not quantifiable. An additional difficulty is the complexity of the tournament, with billions of different outcomes, making it very difficult to obtain accurate guesses of the probabilities of certain events. In the particular case of the African championship it is still more unclear to estimate the strengths of the participating teams or even to determine the divergence of the teams' strengths, since many teams or players are not so well-known as the teams from Europe or South America. Hence, the focus of this article is not to make an exact forecast, which seems not reasonable due to many unpredictable events, but to make the discrepancy between the participating teams \textit{quantifiable} and to measure the chances of each team. This approach is underlined by the fact that supporters of the participating teams typically study the tournament structure after the group draw in order to figure out whether their teams have a rather simple or hard way to the final. Hence, the aim is to quantify the difficulty for each team to proceed to the different stages of the tournament.

\subsection{State of the art}

We give some background on modelling football matches. A series of statistical models have been proposed in the literature for the prediction of football outcomes. They can be divided into two broad categories. The  first one, the result-based model, models  directly the probability of a game outcome (win/loss/draw), while the second one, the score-based model, focusses on the prediction of the exact match score. In this article the second approach is used since the match score is a non-neglecting, very important factor in the group phase of the championship and it also implies a model for the first one. In contrast to the FIFA World Cup, where the two best teams in each group of the preliminary round qualify for the round of $16$, the situation becomes more difficult in the Africa Cup of Nations 2019, where also the four best third-placed teams in the group phase qualify for the round of $16$. As we have seen in former World Cups before 1994 or during the European Championship 2016, in most cases the goal difference is the crucial criterion which decides whether a third-placed team moves on to the round of 16 or is eliminated in the preliminary round. This underlines the importance and necessity of estimating the exact score of each single match and not only the outcome (win/loss/draw).

There are several models for this purpose and most of them involve a Poisson model.  The easiest model,  \cite{Le:97}, assumes independence of the goals scored by each team and that each score can be modeled by a Poisson regression model.  Bivariate Poisson models  were proposed earlier by \cite{Ma:82} and extended by  \cite{DiCo:97} and  \cite{KaNt:03}. A short overview on different Poisson models  and related models like generalised Poisson models or zero-inflated models are given in \cite{ZeKlJa:08} and  \cite{ChSt:11}. 
 Possible covariates for the above models may be divided into two major categories: those  containing ``prospective'' informations and those containing ``retrospective'' informations. The first category contains other forecasts, especially bookmakers' odds, see e.g.  \cite{LeZeHo:10a}, \cite{LeZeHo:12} and references therein. This approach relies on the fact that bookmakers have a strong economic incentive to rate the result correctly and that they can be seen as experts in the matter of the forecast of sport events. However,  their forecast models remain undisclosed and rely on  information that is not publicly available.  
The second category contains only historical data and no other forecasts.
Models based on the second category allow to explicitly model the influence of the covariates (in particular, attack/defense strength/weakness). Therefore,   this approach is pursued using a Poisson regression model for the outcome of single matches.  

Since the  Africa Cup of Nations 2019 is a more complex tournament, involving for instance effects such as group draws, e.g. see  \cite{De:11}, and dependences of the different matches,   Monte-Carlo simulations  are used to forecast the whole course of the tournament. For a more detailed summary on statistical modeling of major international football events, see \cite{GrScTu:15}  and references therein. 

Different similar models based on  Poisson regression of increasing complexity (including discussion, goodness of fit and comparing them in terms of scoring functions) were analysed and used in \cite{gilch-mueller:18} for the prediction of the FIFA World Cup 2018. Among the models therein, in this article we will make use of the most promising Poisson model and omit further comparison and validation of different (similar) models. The model under consideration will not only use for estimating the teams' chances to win the Africa Cup but also to answer questions like how the possible qualification of third-ranked teams in the group phase affects the chances of the top favourites.
 Moreover, since the tournament structure of the Africa Cup of Nations 2019 has changed in this edition to 24 participating teams, a comparison with previous editions of this tournament seems to be quite difficult due to the heavy influence of possible qualifiers for the round of 16 as  third-ranked teams. 

Finally, let me say some words on the data available for feeding our regression model.
These days a lot of data on possible  covariates for  forecast models is available.  \cite{GrScTu:15} performed a variable selection on various covariates and found that the three most significant retrospective covariates are the FIFA ranking followed by the number of Champions league and Euro league players of a team. In this article the Elo ranking (see \texttt{http://en.wikipedia.org/wiki/World\_Football\_Elo\_Ratings}) is preferably considered instead of the FIFA ranking (which is a simplified Elo ranking since July 2018), since the calculation of the FIFA ranking changed over time and the Elo ranking is  more widely used in football forecast models. See also \cite{GaRo:16} for a discussion on this topic and a justification of the Elo ranking. At the time of this analysis the composition and the line ups of the teams have not been announced and hence the two other covariates are not available.
This is  one of the reasons that the model under consideration is solely based  on the Elo points and matches  of the participating teams on neutral ground since 2010. The obtained results show that, despite the simplicity of the model, the model under consideration shows a good fit, the obtained forecast is conclusive and give \textit{quantitative insights} in each team's chances. In particular, we quantify the chances of each team to proceed to a specific phase of the tournament, which allows also to compare the challenge for each team to proceed to the final.

\subsection{Questions under consideration}
\label{subsec:goals}
The simulation in this article works as follows: each single match is modeled as $G_{A}$:$G_{B}$, where $G_{A}$ (resp.~$G_{B}$) is the number of goals scored by team A (resp.~by team B).  So much the worse not only a single match is forecasted but the course of the whole tournament. Even the most probable tournament outcome has a probability, very close to zero  to be actually realized. Hence, deviations of the true tournament outcome from the model's most probable one are not only possible, but most likely. However, simulations of the tournament yield estimates of the  probabilities for each team to reach different stages of the tournament and allow to make the different team's chances \textit{quantifiable}. In particular, we are interested to give quantitative insights into the following questions:
\begin{enumerate}
\item How are the probabilities that a team wins its group or will be eliminated in the group stage?
\item Which team has the best chances to become new African champion?
\item What is the effect of the fact that the four best third-ranked teams in the group phase qualify for the round of 16? How does it affect the chances of the top favourites?
\end{enumerate}
As we will see, the model under consideration in this article  favors Senegal (followed by Nigeria) to win the Africa Cup of Nations 2019.


\section{The model}
\label{sec:model}

\subsection{Involved data}
The model used in this article was proposed in \cite{gilch-mueller:18} (together with several similar bi-variate Poisson models) as \textit{Nested Poisson Regression} and is based on the World Football Elo ratings of the teams. It is based on the Elo rating system, see  \cite{Elo:78}, but includes modifications to take various football-specific variables (like home advantage, goal difference, etc.) into account. The Elo ranking is published by the website \texttt{eloratings.net}.
 The Elo ratings as they were  on 12 april 2019  for the top $5$ participating nations (in this rating) are as follows:
\begin{center}
\begin{tabular}{|c|c|c|c|c|}
\hline
Senegal & Nigeria & Morocco & Tunisia & Ghana \cr
\hline
1764 & 	1717    & 	1706 & 1642	& 1634 \cr
\hline
\end{tabular}
\end{center}
The forecast  of the outcome of a match between teams $A$ and $B$ is modelled as 
$$
G_A\ : \ G_B,
$$
where $G_A$ (resp. $G_{B}$) is the number of goals scored by team $A$ (resp. $B$).
The  model is   based on  a Poisson regression model, where we assume  $(G_{A}, G_{B})$ to be a bivariate Poisson distributed random variable;
see  \cite[Section 8]{gilch-mueller:18} for a discussion on other underlying distributions for $G_A$ and $G_B$. The distribution of $(G_{A}, G_{B})$   
will depend on the current Elo ranking $\elo{A}$ of team $A$ and Elo ranking $\elo{B}$ of team $B$. The model is  fitted  
using  all matches of Africa Cup of Nations 2019 participating teams on \textit{neutral} playground between 1.1.2010 and 12.04.2019. Matches, where one team plays at home, have usually a drift towards the home team's chances, which we want to eliminate. In average, we have for each team $29$ matches from the past and for the top teams even more. 
In the following subsection we explain the model for forecasting a single match, which in turn is used for simulating the whole tournament  and  determining the likelihood of the success for each participant. 

\subsection{Nested Poisson regression}
We now present a \textit{dependent} Poisson regression approach which will be the base for the whole simulation. The number of goals $G_A$, $G_B$ respectively, shall be a Poisson-distributed random variable with rate $\lambda_{A|B}$, $\lambda_{B|A}$ respectively. As we will see one of the rates (that is, the rate of the weaker team) will depend on the concrete realisation of the other random variable (that is, the simulated number of scored goals of the stronger team).
\par
In the following we will always assume that $A$ has \textit{higher} Elo score than $B$. This assumption can be justified, since usually the better team dominates the weaker team's tactics. Moreover the number of goals the stronger team scores has an impact on the number of goals of the weaker team. For example,  if team $A$ scores  $5$ goals it is more likely that $B$ scores also $1$ or $2$ goals, because the defense of team $A$ lacks in concentration  due to the expected victory. If the stronger team $A$ scores only $1$ goal, it is more likely that $B$ scores no or just one goal, since team $A$ focusses more on the defence  and secures the victory.
\par
The Poisson rates $\lambda_{A|B}$ and $\lambda_{B|A}$ are now determined as follows: 
\begin{enumerate}
\item In the first step we model the number of goals $\tilde G_{A}$ scored  by team $A$ only in dependence of the opponent's Elo score $\elo{}=\elo{B}$. The random variable $\tilde G_{A}$  is modeled as a Poisson distribution with parameter $\mu_{A}$. The parameter $\mu_{A}$ as a function of the Elo rating $\elo{\O}$ of the opponent $\O$ is given as
\begin{equation}\label{equ:independent-regression1}
\log \mu_A(\elo{\O}) = \alpha_0 + \alpha_1 \cdot \elo{\O},
\end{equation}
where $\alpha_0$ and $\alpha_1$ are obtained via Poisson regression.  
\item Teams of similar Elo scores  may have different strengths in attack and defense. To take this effect into account  we model the  number of goals team $B$ receives  against a team of Elo score  $\elo{}=\elo{A}$ using a Poisson distribution with parameter $\nu_{B}$. The parameter $\nu_{B}$ as a function of the Elo rating $\elo{\O}$ is given as
\begin{equation}\label{equ:independent-regression2}
\log \nu_B(\elo{\O}) = \beta_0 + \beta_1 \cdot \elo{\O},
\end{equation}
where the parameters $\beta_0$ and $\beta_1$ are obtained via Poisson regression.
\item Team $A$  shall in average score $\mu_A\bigr(\elo{B}\bigr)$ goals against team $B$, but team $B$ shall have $\nu_B\bigl(\elo{A}\bigr)$ goals against. As these two values rarely coincides we model the numbers of goals $G_A$ as a Poisson distribution with parameter
$$
\lambda_{A|B} = \frac{\mu_A\bigl(\elo{B}\bigr)+\nu_B\bigl(\elo{A}\bigr)}{2}.
$$
\item The number of goals $G_B$ scored by $B$ is assumed to depend on the Elo score $E_A=\elo{A}$ and additionally on the outcome of $G_A$. More precisely, $G_B$ is modeled as a Poisson distribution with parameter $\lambda_B(E_A,G_A)$ satisfying
\begin{equation}\label{equ:nested-regression1}
\log \lambda_B(E_A,G_A) = \gamma_0 + \gamma_1 \cdot E_A+\gamma_2 \cdot G_A.
\end{equation}
The parameters $\gamma_0,\gamma_1,\gamma_2$ are obtained by Poisson regression. Hence,
$$
\lambda_{B|A} = \lambda_B(E_A,G_A).
$$
\item The result of the match $A$ versus $B$ is simulated by realizing $G_A$ first and then  realizing $G_B$ in dependence of the realization of $G_A$. 
\end{enumerate}
For a better understanding, we give an example and consider the match Senegal vs. Ivory Coast: Senegal has $1764$ Elo points while Ivory Coast has $1612$ points. Against a team of Elo score $1612$ Senegal is assumed to score in average
$$
\mu_{\textrm{Senegal}}(1612)=\exp(2.73   -0.00145\cdot 1612)=1.48
$$
goals, while Ivory Coast receives against a team of Elo score $1764$ in average
$$
\mu_{\textrm{Ivory Coast}}(1764)=\exp(-4.0158   + 0.00243\cdot 1764)=1.31
$$
goals. Hence, the number of goals, which Senegal will score against Ivory Coast, will be modelled as a Poisson distributed random variable with rate
$$
\lambda_{\textrm{Senegal}|\textrm{Ivory Coast}}=\frac{1.48+1.31}{2}=1.395.
$$
The average number of goals, which Ivory Coast scores against a team of Elo score $1764$ provided that $G_A$ goals against are received, is modelled by a Poisson random variable with rate
$$
\lambda_{\textrm{Ivory Coast}|\textrm{Senegal}}=\exp(1.431 -0.000728\cdot 1764+  0.137\cdot G_A);
$$
e.g., if $G_A=1$ then $\lambda_{\textrm{Ivory Coast}|\textrm{Senegal}}=1.33$.
\par
As a final remark, let me mention that the presented dependent approach may also be justified through the definition of  conditional probabilities:
$$
\mathbb{P}[G_A=i,G_B=j] = \mathbb{P}[G_A=i]\cdot \mathbb{P}[G_B=j \mid G_A=i] \quad \forall i,j\in\mathbb{N}_0.
$$
For comparision of this model in contrast to similar Poisson models, we refer once again to \cite{gilch-mueller:18}. In the following subsections we present some regression plots and will test the goodness of fit.

\subsection{Regression plots}

\begin{figure}[ht]
\begin{center}
\includegraphics[width=6cm]{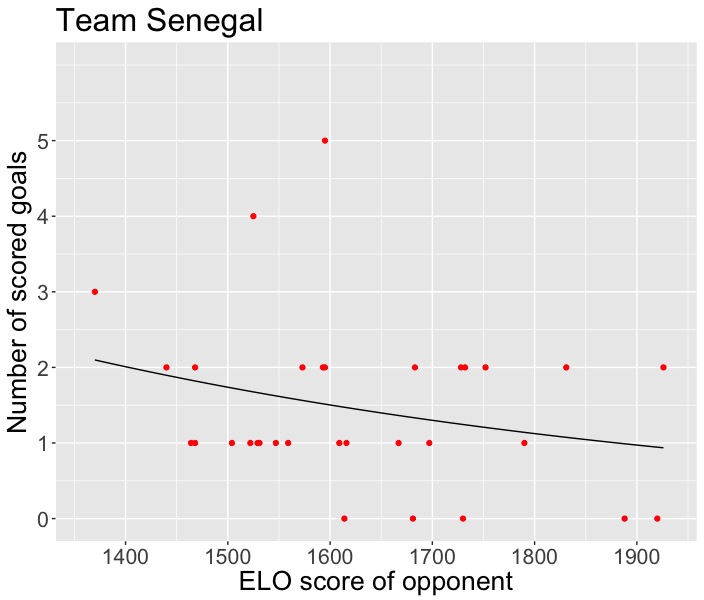}
\hfill
\includegraphics[width=6cm]{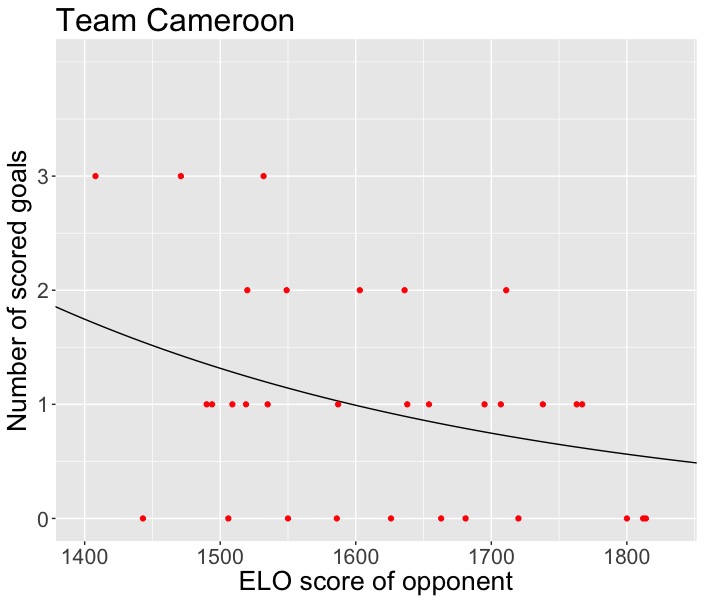} 
\end{center}
\caption{Plots for the number of goals scored by Senegal and Cameroon in regression \eqref{equ:independent-regression1}.}
\label{fig:regression-plot-attack}
\end{figure}

As two examples of interest, we sketch in Figure \ref{fig:regression-plot-attack} the results of the regression in  (\ref{equ:independent-regression1}) for the number of goals scored by Senegal and Cameroon. The  dots show the observed data (i.e, number of scored goals on the $y$-axis in dependence of the opponent's strength on the $x$-axis) and the  line is the estimated mean $\mu_A$ depending on the opponent's Elo strength.

\begin{figure}
\begin{center}
\includegraphics[width=6cm]{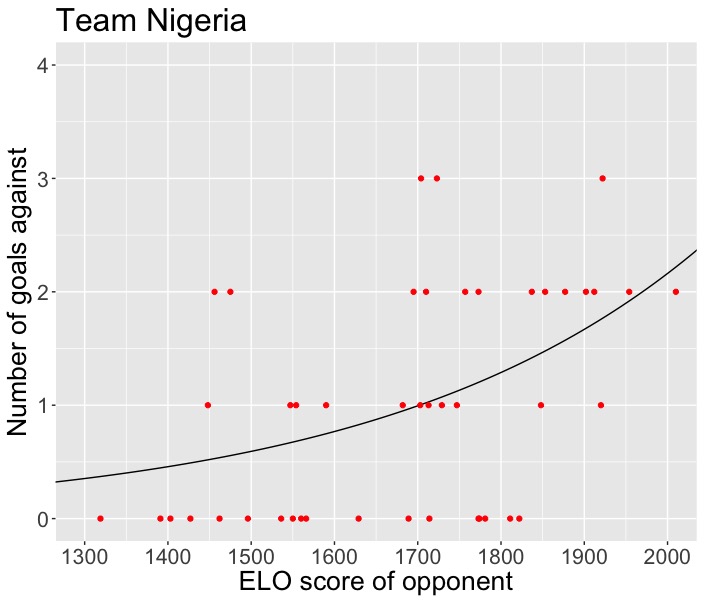}
\hfill
\includegraphics[width=6cm]{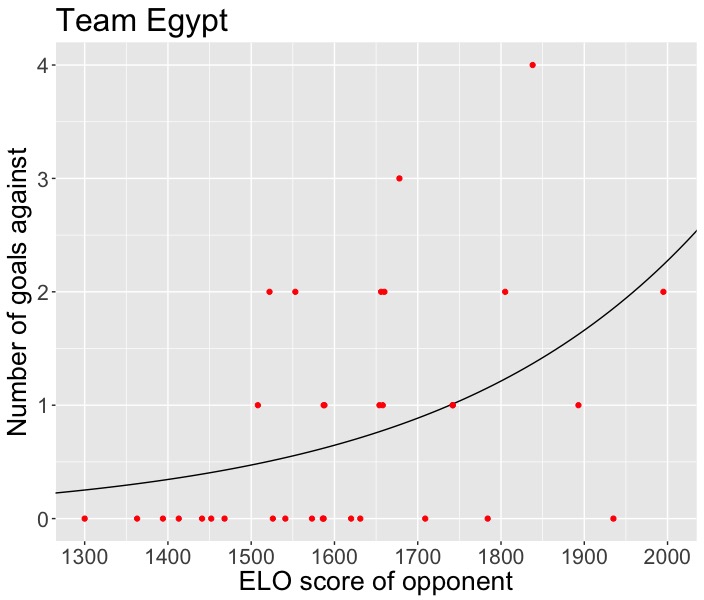} 
\end{center}
\caption{Plots for the number of goals against for  Nigeria and Egypt in regression \eqref{equ:independent-regression2}.}
\label{fig:regression-plot-defense}
\end{figure}

Analogously,   Figure \ref{fig:regression-plot-defense} sketches  the regression in  (\ref{equ:independent-regression2}) for the (unconditioned) number of goals against of Nigeria and Egypt in dependence of the opponent's Elo ranking. The  dots show the observed data (i.e., the number of goals against in the matches from the past) and the  line is the estimated mean $\nu_B$ for the number of goals against.

\subsection{Goodness of fit tests}\label{subsubsection:gof}
We check goodness of fit of  the Poisson regressions in (\ref{equ:independent-regression1}) and (\ref{equ:independent-regression2})  for all participating teams. For each team $\mathbf{T}$ we calculate the following  $\chi^{2}$-statistic from the list of matches from the past:
$$
\chi_\mathbf{T} = \sum_{i=1}^{n_\mathbf{T}} \frac{(x_i-\hat\mu_i)^2}{\hat\mu_i},
$$
where $n_\mathbf{T}$ is the number of matches of team $\mathbf{T}$, $x_i$ is the number of scored goals of team $\mathbf{T}$ in match $i$ and $\hat\mu_i$ is the estimated Poisson regression mean in dependence of the opponent's historical Elo points. 
\par
We observe that most of the teams have a very good fit, except Namibia with a $p$-value of $0.048$. In average, we have a $p$-value of $0.476$. In Table \ref{table:godness-of-fit} the $p$-values for some of the top teams are given. 
\begin{table}[H]
\centering
\begin{tabular}{|l|c|c|c|c|c|}
  \hline
 Team & Senegal & Nigeria & Egypt & Ivory Coast & South Africa 
 \\ 
  \hline
  $p$-value & 0.74 &0.10 &  0.60 & 0.94  &0.72
    \\
   \hline
\end{tabular}
\caption{Goodness of fit test for the  Poisson regression   in  (\ref{equ:independent-regression1}) for some of the top  teams. }
\label{table:godness-of-fit}
\end{table}
Similarly, we can calculate a $\chi^{2}$-statistic for each team which measures the goodness of fit for the regression in  (\ref{equ:independent-regression2}) which models the number of goals against. Here, we  get an average $p$-value of $0.67$; see  Table \ref{table:godness-of-fit2}. 
\begin{table}[H]
\centering
\begin{tabular}{|l|c|c|c|c|c|}
  \hline
 Team & Senegal & Nigeria & Egypt & Ivory Coast & South Africa 
 \\ 
  \hline
  $p$-value & 0.99 &0.79 &  0.38 & 0.51  &0.76
    \\
   \hline
\end{tabular}
\caption{Goodness of fit test for the  Poisson regression   in  (\ref{equ:independent-regression2}) for some of the top  teams. }
\label{table:godness-of-fit2}
\end{table}
Finally, we test the goodness of fit for the regression in (\ref{equ:nested-regression1}) which models the number of goals against of the weaker team in dependence of the number of goals which are scored by the stronger team. We obtain an average $p$-value of $0.33$; see Table \ref{table:godness-of-fit3}. As a conclusion, the $p$-values suggest good fits.
\begin{table}[H]
\centering
\begin{tabular}{|l|c|c|c|c|c|}
  \hline
 Team & Senegal & Nigeria & Egypt & Ivory Coast & South Africa 
 \\ 
  \hline
  $p$-value & 0.99 &0.38 &  0.27 & 0.78  &0.74
    \\
   \hline
\end{tabular}
\caption{Goodness of fit test for the  Poisson regression   in  (\ref{equ:nested-regression1}) for some of the top  teams. }
\label{table:godness-of-fit3}
\end{table}


\subsection{Deviance analysis}
We calculate the null and residual deviances for each team for the regressions in (\ref{equ:independent-regression1}), (\ref{equ:independent-regression2}) and (\ref{equ:nested-regression1}). Tables \ref{table:deviance-IndPR1}, \ref{table:deviance-IndPR2} and \ref{table:deviance-NPR1} show the deviance values and the $p$-values for the residual deviance for some of the top teams.  Most of the $p$-values are not low, except for Nigeria. We remark that the level of significance of the covariates is also of fluctuating quality, but it is still reasonable in many cases.
\begin{table}[ht]
\centering
\begin{tabular}{|l|c|c|c|}
  \hline
 Team & Null deviance & Residual deviance & $p$-value 
 \\ 
  \hline
 Senegal & 28.14 &         26.34 &                   0.66 \\ 
 Nigeria & 71.36 &  66.39 &   0.03\\  
   Egypt & 43.94 &         38.15 &                     0.29\\ 
    Cote d'Ivoire & 47.15    &     46.8&                    0.71\\ 
   South Africa & 12.0 &          10.49 &                    0.65 \\ 
   \hline
\end{tabular}
\caption{Deviance analysis for some top  teams in  regression   (\ref{equ:independent-regression1})}
\label{table:deviance-IndPR1}
\end{table}

\begin{table}[h]
\centering
\begin{tabular}{|l|c|c|c|}
  \hline
  Team & Null deviance & Residual deviance & $p$-value
  \\ 
  \hline
 Senegal & 19.41 &         19.21 &                   0.94\\ 
 Nigeria & 58.50 &        45.35 &                    0.87\\ 
 Egypt & 49.63 & 38.09 & 0.29 \\ 
   Cote d'Ivoire & 69.97 & 59.61 & 0.25 \\ 
   South Africa & 12.14 &      11.92 &               0.53\\ 
     \hline
\end{tabular}
\caption{Deviance analysis for some top  teams in  regression  (\ref{equ:independent-regression2})}
\label{table:deviance-IndPR2}
\end{table}
\begin{table}[h]
\centering
\begin{tabular}{|l|c|c|c|}
  \hline
  Team & Null deviance & Residual deviance & $p$-value
  \\ 
  \hline
 Senegal & 28.1 &         24.8 &                   0.69\\ 
 Nigeria & 71.4 & 62.1         &                    0.05\\ 
 Egypt & 43.94 & 37.98 & 0.25 \\ 
   Cote d'Ivoire & 47.15 & 45.45 & 0.73 \\ 
   South Africa & 12.01 &  10.36     &               0.58\\ 
     \hline
\end{tabular}
\caption{Deviance analysis for some top teams in  regression  (\ref{equ:nested-regression1})}
\label{table:deviance-NPR1}
\end{table}

\section{Africa Cup of Nations 2019 Simulations}
\label{sec:simulation}

Finally, we come to the simulation of the Africa Cup of Nations 2019, which allows us to answer the questions formulated in Section \ref{subsec:goals}. We simulate each single match of the Africa Cup of Nations 2019 according to the model presented in Section \ref{sec:model}, which in turn allows us to simulate the whole Africa Cup tournament. After each simulated match we update the Elo ranking according to the simulation results. This honours teams, which are in a good shape during a tournament and perform maybe better than expected.
Overall, we perform  $100.000$ simulations of the whole tournament, where we reset the Elo ranking at the beginning of each single tournament simulation.

\subsection{Single Matches}
\begin{figure}[ht]
\begin{center}
\includegraphics[width=15cm]{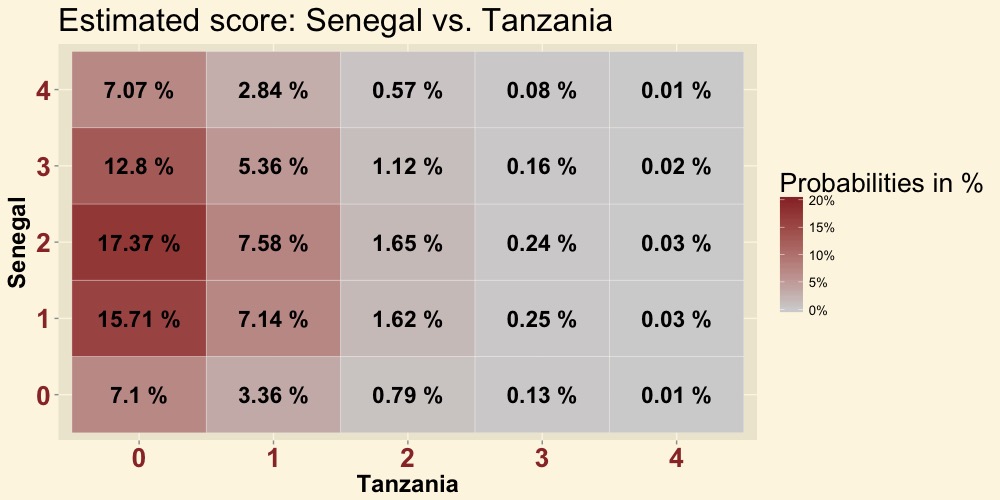}
\end{center}
\caption{Probabilities for the score of the match Senegal vs. Tanzania in Group C.}
\label{table:SN-TZ}
\end{figure}

As the basic element of our simulation is the simulation of single matches, we visualise how to quantify the outcomes of single matches. Group C starts with the match between Senegal and Tanzania. According to our model we have the probabilities presented in Figure \ref{table:SN-TZ} for the result of this match: the most probable score is a $2-0$ victory of Senegal, but a $3-0$ or $1-0$ win would also be among the most probable scores.

\subsection{Group Forecast} 

Among football experts and fans a first natural question after the group draw is to ask how likely it is that the different teams survive the group stage and move on to the round of $16$. Since the individual teams' strength and weaknesses are rather hard to quantify in the sense of tight facts, one of our main aims is to quantify the chances for each participating team to proceed to the round of $16$. With our model we are able to quantify the chances in terms of probabilities how the teams will end up in the group stage. In the following tables \ref{tab:groupA}-\ref{tab:groupF} we present these probabilities obtained from our simulation, where we give the probabilities of winning the group, becoming runner-up, to qualify as one the best third-placed teams or to be eliminated in the group stage. In Group D, the toughest group of all, a head-to-head fight between Morocco, Ivory Coast and South Africa is expected with slight advantage for the team from Ivory Coast.

\begin{table}[ht]
\centering
\begin{tabular}{|r|cccc|}
  \hline
 Team &  1st & 2nd & Qualified as Third & Preliminary Round \\ 
  \hline
Egypt & 51.00 & 28.30 & 11.30 & 9.50 \\ 
DR of Congo & 32.00 & 31.80 & 15.60 & 20.70 \\ 
Uganda  & 4.70 & 14.10 & 16.00 & 65.10 \\ 
Zimbabwe  & 12.40 & 25.80 & 21.20 & 40.60 \\
   \hline
\end{tabular}
\caption{Probabilities for Group A}
\label{tab:groupA}
\end{table}

\begin{table}[ht]
\centering
\begin{tabular}{|r|cccc|}
  \hline
 Team &  1st & 2nd & Qualified as Third & Preliminary Round \\ 
  \hline
 Nigeria  & 53.90 & 26.90 & 10.90 & 8.40 \\ 
Guinea  & 25.80 & 31.70 & 17.20 & 25.40 \\ 
 Madagascar  & 16.10 & 25.90 & 20.50 & 37.60 \\ 
 Burundi  & 4.30 & 15.60 & 17.20 & 62.90 \\
   \hline
\end{tabular}
\caption{Probabilities for Group B}
\label{tab:groupB}
\end{table}
\begin{table}[ht]
\centering
\begin{tabular}{|r|cccc|}
  \hline
 Team &  1st & 2nd & Qualified as Third & Preliminary Round \\ 
  \hline
Senegal & 54.40 & 27.80 & 10.80 & 7.10 \\ 
Algeria & 28.50 & 31.90 & 17.40 & 22.10 \\ 
Kenya & 12.30 & 24.80 & 21.20 & 41.70 \\ 
Tanzania & 4.80 & 15.50 & 16.70 & 63.10 \\
   \hline
\end{tabular}
\caption{Probabilities for Group C}
\label{tab:groupC}
\end{table}

\begin{table}[ht]
\centering
\begin{tabular}{|r|cccc|}
  \hline
 Team &  1st & 2nd & Qualified as Third & Preliminary Round \\ 
  \hline
Morocco  & 29.40 & 27.10 & 17.50 & 26.00 \\
Ivory Coast & 33.60 & 28.80 & 16.70 & 20.90 \\ 
South Africa & 30.40 & 29.00 & 17.20 & 23.40 \\ 
Namibia  & 6.60 & 15.10 & 17.40 & 60.90 \\
   \hline
\end{tabular}
\caption{Probabilities for Group D}
\label{tab:groupD}
\end{table}

\begin{table}[H]
\centering
\begin{tabular}{|r|cccc|}
  \hline
 Team &  1st & 2nd & Qualified as Third & Preliminary Round \\ 
  \hline
Tunisia & 49.60 & 28.60 & 13.50 & 8.30 \\ 
Mali  & 32.10 & 37.50 & 19.00 & 11.40 \\ 
Mauritania & 4.10 & 9.10 & 11.40 & 75.40 \\ 
Angola & 14.30 & 24.80 & 27.00 & 33.90 \\
   \hline
\end{tabular}
\caption{Probabilities for Group E}
\label{tab:groupE}
\end{table}

\begin{table}[H]
\centering
\begin{tabular}{|r|cccc|}
  \hline
 Team &  1st & 2nd & Qualified as Third & Preliminary Round \\ 
  \hline
  Cameroon & 38.80 & 42.60 & 11.90 & 6.80 \\ 
  Ghana & 55.70 & 32.00 & 7.90 & 4.40 \\ 
    Benin & 4.60 & 19.70 & 33.70 & 42.00 \\   
   Guinea-Bissau & 0.90 & 5.70 & 11.00 & 82.30 \\
   \hline
\end{tabular}
\caption{Probabilities for Group F}
\label{tab:groupF}
\end{table}

\subsection{Playoff Round Forecasts}

Finally, according to our simulations we summarise the probabilities for each team to win the tournament, to reach certain stages of the tournament or to qualify for the round of last 16 as one of the best thirds. The result is  presented in Table  \ref{tab:nested18}. E.g., Senegal will at least reach the quarterfinals with a probability of $67,70\%$, while Ghana has a $17\%$ chance to reach the final.
 The regression model  favors Senegal, followed by Nigeria, Ivory Coast and Egypt, to become new football champion of Africa. 

\begin{table}[H]
\centering
\begin{tabular}{|r|ccccc|}
  \hline
 Team & Champion & Final & Semifinal & Quarterfinal & Last16 \\ 
  \hline
 Senegal & 15.40 & 25.20 & 41.20 & 67.70 & 92.90 \\ 
   Nigeria & 12.10 & 22.70 & 37.30 & 59.90 & 91.60 \\ 
   Ivory Coast & 10.20 & 17.70 & 31.10 & 51.90 & 79.10 \\ 
   Egypt & 10.10 & 19.20 & 34.60 & 56.60 & 90.60 \\ 
   Ghana & 8.60 & 17.00 & 30.50 & 57.20 & 95.40 \\ 
   South Africa & 8.40 & 15.50 & 28.50 & 48.80 & 76.50 \\ 
   Morocco & 8.30 & 15.30 & 28.20 & 48.20 & 73.90 \\ 
   Tunisia & 5.80 & 11.90 & 23.20 & 45.50 & 91.70 \\ 
   Algeria & 5.10 & 10.30 & 21.40 & 43.30 & 77.80 \\ 
   Guinea & 3.40 & 8.10 & 17.90 & 37.60 & 74.60 \\ 
   Cameroon & 3.00 & 9.00 & 22.30 & 50.70 & 93.30 \\ 
   DR Congo & 3.00 & 7.70 & 19.00 & 40.00 & 79.10 \\ 
   Mali & 1.60 & 5.00 & 13.20 & 32.70 & 88.50 \\ 
   Madagascar & 1.60 & 4.10 & 10.50 & 25.40 & 62.40 \\ 
   Kenya & 1.10 & 3.10 & 9.10 & 23.90 & 58.40 \\ 
   Angola & 1.00 & 2.80 & 8.00 & 22.10 & 66.10 \\ 
   Zimbabwe & 0.40 & 1.80 & 7.40 & 22.80 & 59.50 \\ 
   Namibia & 0.30 & 1.20 & 4.20 & 13.20 & 39.10 \\ 
   Uganda & 0.10 & 0.50 & 2.60 & 10.30 & 34.90 \\ 
   Tanzania & 0.10 & 0.50 & 2.60 & 10.10 & 36.90 \\ 
   Mauritania & 0.10 & 0.40 & 1.50 & 5.90 & 24.40 \\ 
   Benin & 0.10 & 0.60 & 3.40 & 15.10 & 58.00 \\ 
   Burundi & 0.00 & 0.20 & 1.60 & 7.90 & 37.00 \\ 
   Guinea-Bissau & 0.00 & 0.00 & 0.30 & 2.60 & 17.60 \\ 
   \hline
\end{tabular}
\caption{Africa Cup of Nations 2019 simulation results for the teams' probabilities to proceed to a certain stage}
\label{tab:nested18}
\end{table}

\subsection{Simulation without third-placed qualifiers}
\label{subsec:withoutThirds}

One important and often asked question is whether the current tournament structure, which allows  third-placed teams in the preliminary round still to qualify for the round of 16, is reasonable or not. In particular, it is the question whether this structure is good or bad for the top teams and to quantify this factor.
Hence, the simulation was adapted in the sense that third-placed teams in the group stage are definitely eliminated, while the winners of those groups, which are intended to play against a third-ranked team in the round of 16, move directly to the quarter finals. This leads to the results in Table \ref{table:ohneDritte}: it shows that the top teams have now slightly higher chances to win the tournament.

\begin{table}[ht]
\centering
\begin{tabular}{|r|ccccc|ccc|}
  \hline
 Team & Champion & Final & 1/2 & 1/4 & Last16 & 1st & 2nd & Pre.Round \\ 
  \hline
Senegal & 15.80 & 25.40 & 43.50 & 74.10 & 82.20 & 54.50 & 27.70 &  17.90 \\ 
  Nigeria & 14.50 & 28.40 & 45.30 & 72.30 & 80.60 & 53.90 & 26.70 &  19.40 \\ 
   Egypt & 11.70 & 22.60 & 41.10 & 67.90 & 79.30 & 50.50 & 28.70 &20.70 \\ 
   Ivory Coast & 9.90 & 17.00 & 30.40 & 51.50 & 62.50 & 33.50 & 29.00 &  37.50 \\ 
   South Africa & 7.90 & 14.40 & 27.40 & 47.90 & 59.60 & 30.80 & 28.70 &  40.50 \\ 
   Ghana & 7.80 & 16.00 & 28.40 & 53.40 & 87.70 & 55.60 & 32.10 &  12.30 \\ 
   Morocco & 7.60 & 13.50 & 26.20 & 45.90 & 56.50 & 29.20 & 27.20 &  43.60 \\ 
   Algeria & 5.10 & 10.20 & 21.90 & 45.90 & 60.10 & 28.50 & 31.70 &  39.80 \\ 
   Tunisia & 4.70 & 9.60 & 19.00 & 39.00 & 77.70 & 49.30 & 28.50 &  22.20 \\ 
   \hline
\end{tabular}
\caption{Adapted Africa Cup of Nations 2019 simulation results, where third-placed teams are definitely eliminated}
\label{table:ohneDritte}
\end{table}

In Table  \ref{table:ohneDritteDifferenzPlayoff} we compare the probabilities of reaching different stages in the case of the adapted tournament (third-ranked teams are definitely eliminated) versus the real tournament structure, which still allows third-ranked teams to qualify for the round of $16$. As one can see, the differences are rather marginal. However, the top favourite teams would profit from the adapted setting slightly. Moreover, many teams have a chance of 10\% or more to qualify for the round of $16$ as one of the best four third-ranked teams. Thus, the chances to win the African championship remain more or less the same, making it neither harder nor easier for top ranked teams to win.


 \begin{table}[H]
\centering
\begin{tabular}{|l|rrrrr|}
  \hline
 Team & Champion & Final & Semifinal & Quarterfinal & Last16 \\ 
  \hline
Senegal & 0.40 & 0.20 & 2.30 & 6.40 & -10.70 \\ 
 Nigeria & 2.40 & 5.70 & 8.00 & 12.40 & -11.00 \\ 
   Egypt & 1.60 & 3.40 & 6.50 & 11.30 & -11.30 \\ 
   Ivory Coast & -0.30 & -0.70 & -0.70 & -0.40 & -16.60 \\ 
   South Africa & -0.50 & -1.10 & -1.10 & -0.90 & -16.90 \\ 
   Ghana & -0.80 & -1.00 & -2.10 & -3.80 & -7.70 \\ 
   Morocco & -0.70 & -1.80 & -2.00 & -2.30 & -17.40 \\ 
   Algeria & 0.00 & -0.10 & 0.50 & 2.60 & -17.70 \\ 
   Tunisia & -1.10 & -2.30 & -4.20 & -6.50 & -14.00 \\ 
   \hline
\end{tabular}
\caption{Difference of probabilities of adapted tournament simulation vs. real tournament structure}
\label{table:ohneDritteDifferenzPlayoff}
\end{table}

\section{Discussion on Related Models}
\label{sec:discussion}

In this section we want to give some quick discussion about the used Poisson models and related models. Of course, the Poisson models we used are not the only natural candidates for modeling football matches. Multiplicative mixtures may lead to overdispersion. Thus, it is desirable to use models having a variance function which is flexible enough to deal with overdispersion and underdispersion. One natural model for this is the \textit{generalised Poisson model}, which was suggested by  \cite{Co:89}. We omit the details
but remark that this distribution has an additional parameter $\varphi$ which allows to model the variance as $\lambda/\varphi^2$; for more details on generalised Poisson regression we refer to  \cite{St:04} and  \cite{Er:06}. Estimations of $\varphi$ by generalised Poisson regression lead to the observation that $\varphi$ is close to $1$ for the most important teams; compare with \cite{gilch-mueller:18}. Therefore, no additional gain is given by the use of the generalised Poisson model.  
\par
Another related candidate for the simulation of football matches is given by the \textit{negative binomial distribution}, where also another parameter comes into play to allow a better fit. However, the same observations as in the case of the generalised Poisson model can be made, that is, the estimates of the  additional parameter lead to a model which is almost just a simple Poisson model. We refer to  \cite{JoZh:09} for a detailed comparison of generalized Poisson distribution and negative Binomial distribution.
\par
For further discussion on adaptions and different models, we refer once again to the discussion section in \cite{gilch-mueller:18}

\section{Conclusion}
A team-specific Poisson regression model for the number of goals in football matches  facing each other in international tournament matches has been used for quantifying the chances of the teams participating in the Africa Cup of Nations 2019. They all include the Elo points of the teams as covariates and use all  matches of the teams since 2010 as underlying data.The fitted model was used for Monte-Carlo simulations  of the Africa Cup of Nations 2019.  According to this simulation, Senegal (followed by Nigeria) turns out to be the top favorite for winning the title. Besides, for every team probabilities of reaching the different stages of the cup are calculated. 

A major part of the statistical novelty of the presented work lies in the construction of the nested regression model. This model  outperforms previous studied models, that use  (inflated) bivariate Poisson regression, when tested on the previous FIFA World Cups 2010, 2014 and 2018; see the technical report \cite{gilch-mueller:18}

\bibliographystyle{apalike}
\bibliography{bib}

\end{document}